\documentclass[a4paper]{article}

\usepackage[english]{babel}
\usepackage[utf8x]{inputenc}
\usepackage[T1]{fontenc}

\usepackage[a4paper,top=3cm,bottom=2cm,left=3cm,right=3cm,marginparwidth=1.75cm]{geometry}
\usepackage{setspace}
\usepackage[table]{xcolor}
\usepackage{amsmath}
\usepackage{mathrsfs}
\usepackage{amssymb, multirow}
\usepackage{enumitem}
\usepackage{graphicx}
\usepackage[colorinlistoftodos]{todonotes}
\usepackage[colorlinks=true, allcolors=blue]{hyperref}
\usepackage{wrapfig}
\usepackage{subcaption}
\usepackage{booktabs,siunitx}
\usepackage{natbib}
\usepackage[linesnumbered,boxruled,lined]{algorithm2e}
\usepackage{algpseudocode}

\usepackage{url}
\newcommand{\proglang}[1]{\textsf{#1}}
\newcommand{\pkg}[1]{\texttt{#1}}

\title{Spatial Covariance Constraints for
Gaussian Mixture Models}
\author{}
\date{}
\author{Hanzhang Lu\footnote{Corresponding Author: University of British Columbia Okanagan Campus, Kelowna, BC, Canada, V1V 1V7. Email: hanzhang.lu@ubc.ca}, Keiran Malott, Venkat Suprabath Bitra, Kirsty Milligan,\\  Sanjeena Subedi, Edana Cassol, Vinita Chauhan, \\ Connor McNairn, Bryan Muir, Prarthana Pasricha,\\ Sangeeta Murugkar, Rowan Thomson, Andrew Jirasek, Jeffrey L.\ Andrews }

\begin{document}
\maketitle

\begin{abstract}

     Although extensive research exists in spatial modeling, few studies have addressed finite mixture model-based clustering methods for spatial data. Finite mixture models, especially Gaussian mixture models, particularly suffer from high dimensionality due to the number of free covariance parameters. This study introduces a spatial covariance constraint for Gaussian mixture models that requires only four free parameters for each component, independent of dimensionality. Using a coordinate system, the spatially constrained Gaussian mixture model enables clustering of multi-way spatial data and inference of spatial patterns. The parameter estimation is conducted by combining the expectation-maximization (EM) algorithm with the generalized least squares (GLS) estimator. Simulation studies and applications to Raman spectroscopy data are provided to demonstrate the proposed model.

\end{abstract} 

\section{Introduction} \label{intro}

In recent years, the structure of data has grown increasingly complex, characterized by a proliferation of information and greater dimensionality. Spatial correlation, a natural phenomenon that refers to the relationship between values and locations of a variable, is common in various datasets. In modeling spatial patterns, well-established methods such as the Poisson point process \citep{streit2010poisson}, Kriging \citep{oliver1990kriging}, and the Matérn covariance function \citep{genton2001classes} have been proposed and developed. However, in the realm of mixture model-based clustering, there is a limited amount of literature. To the best of our knowledge, the work presented by \cite{lee2025clustering} is the sole contribution that integrates a mixture of regression models with a Markov random field to cluster spatial data. 

Moreover, in the real world, space is typically considered to be two-dimensional or three-dimensional, so spatial data is naturally arranged in two- or three-dimensional order. The intrinsic structure of spatial data also makes it a part of multi-way data analysis, which has garnered considerable attention in recent years. In terms of the model-based clustering, many matrix variate approaches \citep{viroli2011finite,dougru2016finite,gallaugher2018finite,tomarchio2022mixtures} and tensor variate techniques \citep{mai2022doubly,tomarchio2023parsimonious} have been proposed. Although the parsimony of these approaches allows for the analysis of relatively high-dimensional data, they are not designed to capture spatial correlations. To foreshadow our motivating real data analysis, consider the images shown in Figure~\ref{fig: examples of tensors two dose levels}. Therein, a clear spatial autocorrelation is present which the previously mentioned matrix-variate models cannot take advantage of in an efficient manner.

To fill this gap, this study introduces a Gaussian mixture model with spatial constraints (SpatGMM) for clustering multi-way spatial data and inferring spatial patterns. Based on the linear spatial correlation model proposed by \cite{worsley1991linear}, a novel covariance structure called the sigmoid decay will be introduced. It assumes that the spatial correlation decreases as the Euclidean distance increases, but in a certain range of distance, the correlation decreases dramatically and then stabilizes. With the Euclidean distance evaluated on a given coordinate system, under the sigmoid decay structure, the covariance matrices can be decomposed as a linear combination of three symmetric matrices. This structure enables the model to estimate spatial covariance matrices using only three parameters, thereby preventing parameter inflation as dimensionality increases. To conduct the parameter estimation for the spatial covariance parameters, a variant of the expectation-maximization (EM) algorithm combined with the generalized least squares (GLS) estimator \citep{browne1974generalized} is used.

\section{Background} \label{sec:back}

Before introducing the relevant background knowledge, we first introduce the notations and the basic definitions. %Because we will alternate between multi-way and multivariate scenarios, we first clarify the notations used in each to avoid confusion. 
In the multi-way data scenario, we denote random matrices and tensors as $\mathscr{X}$ and the corresponding realization as $\mathcal{X}$. In the multivariate case, the random vector is denoted as $\mathbf{X}$, with $\mathbf{x}$ representing a specific realization. Regarding the scalars, the random variable is denoted by $X$. Furthermore, suppose we have an $M$-order random tensor $\mathscr{X} \in \mathbb{R}^{p_1\times\dots\times p_M}$ along with a coordinate system $\mathcal{C}$, then the total number of variables can be found as  $p = \prod_{m=1}^M p_m$. The vectorization operator $\operatorname{vec}(\cdot)$ transform a tensor $\mathscr{X}$ to a vector $\mathbf{X} = \operatorname{vec}(\mathscr{X}) \in \mathbb{R}^{p\times 1}$, with $X_{i1, \dots, iM}$ being its $j$th element, where $j=1+\sum_{m=1}^M(i_m-1)\prod_{m^\prime = 1}^{m-1}p_{m^\prime}$. For an element $X_{i_1,\dots,i_m}$ in $\mathscr{X}$, or correspondingly $X_j$ in $\mathbf{X}$, the coordinate can be its subscript $(i_1,\dots,i_m)$ or a one-to-one projection vector $\mathbf{c}_j$ of $(i_1,\dots,i_M)$. The coordinate system $\mathcal{C}$ is a set containing the unique corresponding coordinates $\mathbf{c}_1,\dots ,\mathbf{c}_p$. For two coordinates $\mathbf{c}_i$ and $\mathbf{c}_j$, the spatial proximity measured by the Euclidean distance between these two elements is calculated via
\begin{equation*}
    d_{ij} = \sqrt{(\mathbf{c}_i-\mathbf{c}_j)^2}.
\end{equation*}
A spatial system is defined as the random vector $\mathbf{X}$ and its coordinate system $\mathcal{C}$.

% \subsection{Notation}
% A spatial system is defined as an $M$-order random tensor $\mathscr{X} \in \mathbb{R}^{p_1\times\dots\times p_M}$ along with a coordinate system $\mathcal{C}$. We define $p = \prod_{m=1}^M p_m$. The vectorization operator $\operatorname{vec}(\cdot)$ transform a tensor $\mathscr{X}$ to a vector $\mathbf{X} = \operatorname{vec}(\mathscr{X}) \in \mathbb{R}^{p\times 1}$, with $X_{i1, \dots, iM}$ being its $j$th element, where $j=1+\sum_{m=1}^M(i_m-1)\prod_{m^\prime = 1}^{m-1}p_{m^\prime}$. For an element $X_{i_1,\dots,i_m}$ in $\mathscr{X}$, or correspondingly $X_j$ in $\mathbf{X}$, the coordinate can be its subscript $(i_1,\dots,i_m)$ or a one-to-one projection vector $\mathbf{c}_j$ of $(i_1,\dots,i_m)$. Hereafter, we will discuss the spatial systems and random variables in terms of vectorization. The coordinate system $\mathcal{C}$ is a set containing the unique corresponding coordinates $\mathbf{c}_1,\dots ,\mathbf{c}_p$. For two coordinates $\mathbf{c}_i$ and $\mathbf{c}_j$, the spatial proximity measured by the Euclidean distance between these two elements is calculated via
% \begin{equation*}
%     d_{ij} = \sqrt{(\mathbf{c}_i-\mathbf{c}_j)^2}.
% \end{equation*}

\subsection{Gaussian Mixture Models} \label{sec: gaussian mixture models}
Model-based clustering is a statistical method that utilizes finite mixture models, first introduced by \cite{wolfe1963object} and subsequently established in the literature \citep{mclachlan2000finite,mcnicholas2016mixture}. A finite mixture model represents the distribution of a heterogeneous population as a combination of several distributions, each corresponding to a homogeneous subpopulation. These subpopulations, known as components, typically follow distributions from a specified family such as Gaussian \citep{scrucca2016mclust}, t-distributions \citep{andrews2012model}, or generalized hyperbolic distributions \citep{browne2015mixture}.

One of the oldest and most classic finite mixture models for clustering is the Gaussian mixture model \citep[GMM;][]{wolfe70}, which serves as the foundation for numerous model-based clustering methods. The density function of a $p$-dimensional, $G$-component GMM is given by
\begin{equation} \label{eq: GMM}
    f(\mathbf{x}\mid \boldsymbol{\vartheta}) = \sum_{g=1}^G \pi_g\phi(\mathbf{x}\mid\boldsymbol{\mu}_g,\boldsymbol{\Sigma}_g),
\end{equation}
where $\phi(\mathbf{x}\mid\boldsymbol{\mu}_g,\boldsymbol{\Sigma}_g)$ is the $g$th $p$-dimensional Gaussian density function with the mean vector $\boldsymbol{\mu}_g$ and the covariance matrix $\boldsymbol{\Sigma}_g$ and $\boldsymbol{\vartheta} = \{\pi_1,\dots,\pi_G, \boldsymbol{\mu}_1,\dots,\boldsymbol{\mu}_G, \boldsymbol{\Sigma}_1, \dots, \boldsymbol{\Sigma}_G\}$ represents the parameter space.

Despite its popularity, the GMM faces several challenges, especially in high-dimensional settings. The primary issue is the large number of free parameters required for the component covariance matrices. For the model in equation \eqref{eq: GMM}, the number of free covariance parameters is calculated as
\begin{equation*}
    \frac{1}{2}G p(p+1),
\end{equation*}
which becomes considerable as the dimensionality $p$ increases. 

\subsection{Constrained Gaussian Models}

To mitigate this issue, researchers have proposed several covariance structures for GMMs that reduce the number of free parameters while better capturing the covariance of data. \cite{celeux95} proposed the Gaussian parsimonious clustering models (GPCM) with an eigen-decomposition of the group covariance matrices,
\begin{equation}
    \boldsymbol{\Sigma}_g = \lambda_g \boldsymbol{\Gamma}_g \boldsymbol{\Omega}_g \boldsymbol{\Gamma}_g^\prime,
\end{equation}
where $\lambda_g = |\boldsymbol{\Sigma}_g|^{1/p}$, $\boldsymbol{\Gamma}_g$ contains the eigenvectors of $\boldsymbol{\Sigma}_g$, and $\boldsymbol{\Omega}_g$ is a diagonal matrix of normalized eigenvalues in decreasing order with $|\boldsymbol{\Omega}_g| = 1$. Imposing constraints on these elements reduces the number of free covariance parameters.

As an extension of the factor analysis model \citep{spearman1987proof}, the mixture of factor analyzers model is presented by \cite{ghahramani1996algorithm}, which assumes $\mathbf{X}_i$ can be expressed as a mixture of linear combinations of $r$-dimensional latent factors $\mathbf{U}_{ig}$, where $r \ll p$, which can be written as
\begin{equation} \label{eq: MFA}
    \mathbf{X}_i = \boldsymbol{\mu}_g + \boldsymbol{\Lambda}_g\mathbf{U}_{ig}+ \boldsymbol{\epsilon}_{ig},
\end{equation}
with probability $\pi_g$, where $g$ is the index of the corresponding component, $\boldsymbol{\mu}_g$ is the $p \times 1$ location vector of $g$th group, $\boldsymbol{\Lambda}_g$ is the $p \times r$ factor loadings of $g$th group, $\mathbf{U}_{ig} \sim \mathcal{N}_r(\mathbf{0},\mathbf{I}_r)$, and $\boldsymbol{\epsilon}_{ig} \sim \mathcal{N}_p(\mathbf{0},\boldsymbol{\Psi}_g)$. Here, $\mathbf{I}_r$ refers to a $r$-dimensional identity matrix, and $\boldsymbol{\Psi}_g = \operatorname{diag}(\psi_{1g},\psi_{2g},\dots,\psi_{pg})$ denotes a $p \times p$ diagonal matrix with elements $(\psi_{1g},\psi_{2g},\dots,\psi_{pg})$. Under \eqref{eq: MFA}, the group covariance can be decomposed as
\begin{equation*}
    \boldsymbol{\Sigma}_g = \boldsymbol{\Lambda}_g\boldsymbol{\Lambda}_g^\prime+\boldsymbol{\Psi}_g.
\end{equation*}
% Consequently, the overall density of $\mathbf{X}_i$ is
% \begin{equation*}
%     f(\mathbf{x}\mid \boldsymbol{\vartheta}) = \sum_{g=1}^G \pi_g\phi(\mathbf{x}\mid\boldsymbol{\mu}_g,\boldsymbol{\Lambda}_g\boldsymbol{\Lambda}_g^\prime+\boldsymbol{\Psi}_g).
% \end{equation*}
The number of free covariance parameters in the MFA is given by
\begin{equation*}
    G\left[pr+p - \frac{1}{2}r(r-1)\right].
\end{equation*}
Hence, the reduction of the total number of free parameters is
\begin{equation*}
    \frac{1}{2}Gp(p+1)-G\left[ pr+p - \frac{1}{2}r(r-1)\right] = \frac{1}{2}G\left[(p-r)^2 - (p+r)\right],
\end{equation*}
which is positive with $(p-r)^2 > (p+r)$. Moreover, by imposing the constraints $\boldsymbol{\Lambda}_g = \boldsymbol{\Lambda}$, $\boldsymbol{\Psi}_g = \boldsymbol{\Psi}$, and $\boldsymbol{\Psi}_g = \boldsymbol{\psi}_g \mathbf{I}$ to allow different components to share covariance parameters, \cite{mcnicholas2008parsimonious} propose a family of eight parsimonious Gaussian model, which reduce the number of free parameters in the covariance structure further.

\subsection{Mixture of Matrix normal distributions} \label{sec: MMN}

Another extension of the GMM is the mixture of matrix normal distributions (MMN) proposed by \cite{viroli2011finite}, which leverages the matrix normal distribution \citep{gupta1992characterization,gupta2018matrix} to accommodate matrix-variate inputs. The matrix normal distribution, also known as the matrix Gaussian distribution, is a special case of the multivariate normal distribution that applies to matrix variate random variables. It features two distinct covariance matrices to capture the variability and correlation among the rows and columns, separately. 

Suppose a $p \times q$ random matrix $\mathscr{X}$ arises from a matrix Gaussian distribution with a $p \times q$ location matrix $\mathbf{M}$, a $p \times p$ across-row covariance matrix $\boldsymbol{\Xi}$ and a $q \times q$ across-column covariance matrix $\boldsymbol{\Omega}$, denoted by $\mathcal{N}_{p \times q}(\mathbf{M},\boldsymbol{\Xi},\boldsymbol{\Omega})$, has the density function, for all $\mathcal{X}\subset \mathscr{X}$, can be written as
\begin{equation} \label{eq: MVG density}
    \phi_{p \times q}(\mathcal{X} \mid \mathbf{M},\boldsymbol{\Xi},\boldsymbol{\Omega}) = \frac{\operatorname{exp}\left\{-\frac{1}{2}\operatorname{tr}\left(\boldsymbol{\Xi}^{-1}(\mathcal{X}-\mathbf{M})\boldsymbol{\Omega}^{-1}(\mathcal{X}-\mathbf{M})^\prime\right)\right\}}{(2\pi)^{\frac{pq}{2}}|\boldsymbol{\Xi}|^{\frac{p}{2}}|\boldsymbol{\Omega}|^{\frac{q}{2}}}.
\end{equation}
One key property of the matrix normal distribution is an equivalent definition that if $\mathscr{X} \sim \mathcal{N}_{p \times q}(\mathbf{M},\boldsymbol{\Xi},\boldsymbol{\Omega})$, then $\operatorname{vec}(\mathscr{X}) \sim \mathcal{N}_{pq}(\operatorname{vec}(\mathbf{M}), \boldsymbol{\Omega}\otimes\boldsymbol{\Xi})$, where $\mathcal{N}_{pq}(\cdot)$ represents the $mp$-dimensional multivariate normal distribution, $\operatorname{vec}(\cdot)$ is the vectorization operator, and $\otimes$ is the Kronecker product. Building on this property, the density of a $G$-component MMN can be written as
\begin{equation} \label{eq: MMN density}
    f(\mathcal{X}\mid \boldsymbol{\vartheta}) = \sum_{g=1}^G \pi_g\phi_{ p \times q}(\mathcal{X} \mid \mathbf{M}_g,\boldsymbol{\Xi}_g,\boldsymbol{\Omega}_g),
\end{equation}
where $\phi_{p \times q}(\mathcal{X} \mid \mathbf{M}_g,\boldsymbol{\Xi}_g,\boldsymbol{\Omega}_g)$ denotes the $g$th matrix-variate Gaussian density function, characterized by the mean matrix $\mathbf{M}_g$, the across-row covariance matrix $\boldsymbol{\Xi}_g$, and the across-column covariance matrix $\boldsymbol{\Omega}_g$. Thus, the MMN can be interpreted as the multivariate GMM with a Kronecker product covariance structure. The Kronecker product decomposes the component covariance matrices into two positive definite matrices, which separately handle the covariance across rows and columns. Besides, this approach also eases the dimensionality challenge associated with the covariance matrices in the multivariate GMM. Specifically, the reduction of the free covariance parameters is
\begin{equation*}
     \frac{1}{2}G \left[pq(pq+1) - p(p+1) - q(q+1)  \right],
\end{equation*}
which is strictly positive.

\subsection{Linear Spatial Covariance Structure} \label{sec: linear spatial covariance structure}
    
The linear spatial covariance structure, or the linear spatial correlation model, proposed by \cite{worsley1991linear} provides an efficient way to model spatial patterns. As a special case of the linear covariance structures discussed by \cite{browne1974generalized}, it assumes a spatial covariance matrix can be decomposed as a linear combination of three predefined matrices. Suppose we have a $p$-dimensional spatial system $\mathbf{X}$ following Gaussian distribution with covariance matrix $\boldsymbol{\Sigma}$ and its coordinate system $\mathcal{C}$, the linear spatial covariance suggests that the covariance 
\begin{equation} \label{eq: quadratic decay}
    \boldsymbol{\Sigma} = \alpha_1 \mathbf{J}-\alpha_2\mathbf{D}+\alpha_3 \mathbf{I},
\end{equation}
where $\boldsymbol{\Sigma}$ is the covariance matrix of $\mathbf{X}$, $\alpha_1, \alpha_2,\alpha_3 \in \mathbb{R}_+$ are the linear spatial parameters, $\mathbf{J}$ is a $p \times p$ matrix of ones, $\mathbf{D}$ is a $p \times p$ matrix of the Euclidean distance evaluated on $\mathcal{C}$, and $\mathbf{I}$ is a $p \times p$ identity matrix. With the distance $d_{ij}$ between the $i$th and $j$th elements $X_i$ and $X_j$, their covariance $\sigma_{ij}$ can be calculated via
\begin{equation} \label{eq: quadratic decay element}
    \sigma_{ij} = \begin{cases}
  \alpha_1 - \alpha_2 d_{ij}^2& \text{ if } i \ne j, \\
  \alpha_1 + \alpha_3 & \text{ if } i=j.
\end{cases}
\end{equation}
The function~\eqref{eq: quadratic decay element} suggests that the spatial covariance decreases quadratically with the Euclidean distance, so this structure is also called the quadratic decay (QD) covariance structure. Moreover, in \eqref{eq: quadratic decay}, except for three predefined matrices, there are only three free parameters. As the dimensionality increases, these three matrices will be expanded, but this will not affect the number of free parameters the QD uses. Hence, the principal advantage of this structure lies in its ability to model a spatial covariance matrix with only three parameters, regardless of dimensionality.

In terms of parameter estimation, \cite{browne1974generalized} presents the generalized least squares (GLS) estimator for estimating the linear covariance structure. Suppose that a set of $p$-dimensional random vectors, denoted as $\mathbf{X}_i$ for $i=1,2,\dots,N$ identically, independently follow $\mathcal{N}(\boldsymbol{\mu},\boldsymbol{\Sigma})$, with a shared coordinate system $\mathcal{C}$. Under the quadratic decay, $\boldsymbol{\Sigma}$ follows \eqref{eq: quadratic decay}. The GLS estimates $\hat{\alpha}_1, \hat{\alpha}_2,\hat{\alpha}_3$ by minimizing the quadratic form of the residual as follows
\begin{equation}\label{eq: residual quadratic form}
    g(\boldsymbol{\alpha}) = \frac{1}{2}(\mathbf{s}-\boldsymbol{\Delta}\boldsymbol{\alpha})^\prime\left(\mathbf{V} \otimes \mathbf{V}\right)^{-1}(\mathbf{s}-\boldsymbol{\Delta}\boldsymbol{\alpha}),
\end{equation}
where $\boldsymbol{\alpha} = (\alpha_1, \alpha_2,\alpha_3)^\prime$ is the vector of the linear spatial covariance parameter, $\mathbf{s} = \operatorname{vec}(\mathbf{S})$ is the vectorization of the sample covariance matrix, $\boldsymbol{\Delta}$ is the design matrix with the columns $\operatorname{vec}(\mathbf{J}), -\operatorname{vec}(\mathbf{D}),\operatorname{vec}(\mathbf{I})$, and $\mathbf{V}$ is the positive definite weight matrix. Since the linear spatial covariance structure is a special case of the linear covariance structure, the GLS estimator can also estimate $\alpha_1, \alpha_2,\alpha_3$ in \eqref{eq: quadratic decay}. There are several desirable properties the GLS estimator possesses, but one worth noting here is that the GLS estimates are also the maximum likelihood estimates (MLE), which are also the minimizers of 
\begin{equation*}
    F = \log |\boldsymbol{\Xi}| - \log |\mathbf{S}| + \operatorname{tr}\left\{\mathbf{S}
    {\boldsymbol{\Xi}}^{-1}\right\} -p.
\end{equation*}

\section{Methodology}\label{method}

\subsection{Covariance Structure} \label{sec: cs}
\begin{figure}[ht]
    \centering
    \includegraphics[width=1\linewidth]{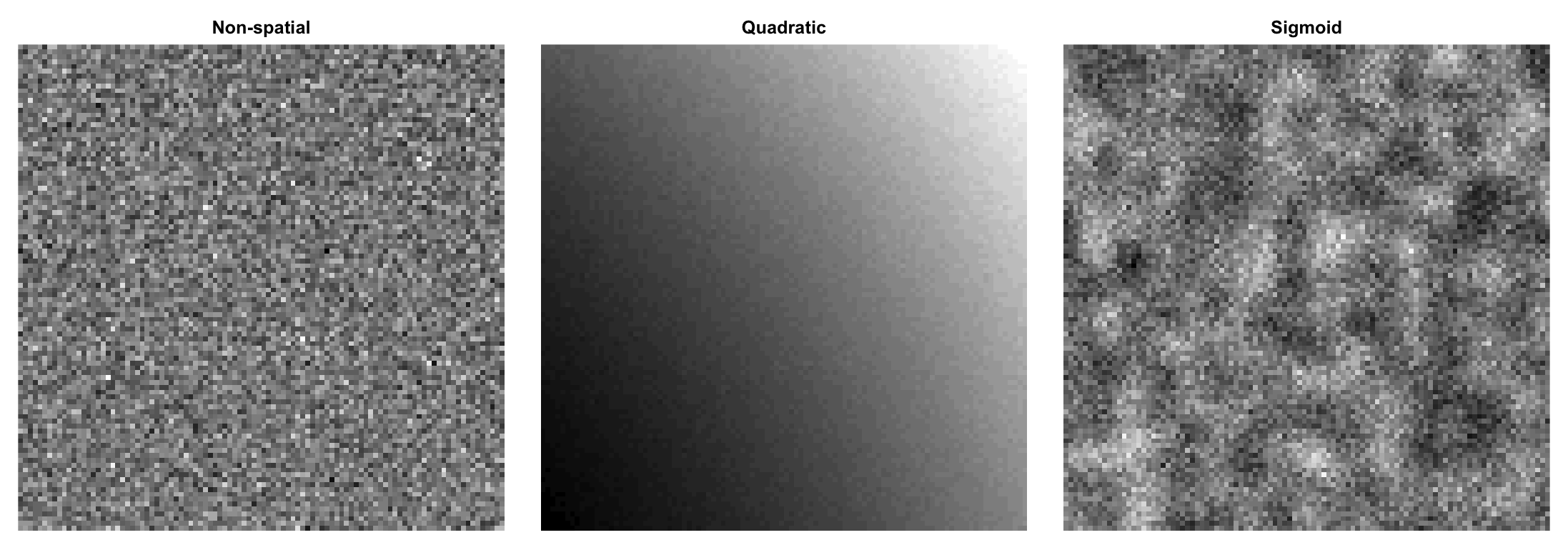}
    \caption{Simulated spatial maps from Gaussian noise (left), quadratic decay (middle), and sigmoid decay (right) models.}
    \label{fig: simmaps}
\end{figure}

Although the QD covariance structure offers us considerable parsimony, it also imposes some restrictions. Since the matrix $\mathbf{D}$ in \eqref{eq: quadratic decay} is entirely determined by the given coordinate system, when facing different coordinate systems, it always requires different normalizations to keep the distance in a reasonable range. Moreover, in the high-dimensional context, collinearity is a significant issue for the QD structure, which renders the matrix non-positive definite. Therefore, to overcome these concerns, we propose a sigmoid decay (SD) structure by changing the quadratic function applied to the distance to a parameterized sigmoid function. It assumes that the spatial correlation diminishes to a specific value at a certain distance and then stabilizes. The introduced sigmoid function in the SD structure is 
\begin{equation} \label{eq: sigmoid function}
    h(x) = a\left( \frac{1}{1+e^{-\beta x + 3}} - s \right),
\end{equation}
where $x \leq 2$, $s = \frac{1}{1+e^{3}}$, $a=(\frac{1}{1+exp(-2\beta+3)}-s)^{-1}$ and $\beta$ is a tuning parameter. It is worth noting that this function always starts at $(0,0)$ and passes through $(2,1)$. Unlike the QD, the distance is always normalized to be in the range of $[0,2]$ regardless of coordinate systems. In this way, the value of the sigmoid parameter $\beta$ is kept in the sensitive and appropriate range. Additionally, this function helps prevent collinearity in the linear covariance structure.
\begin{figure}[!ht]
    \centering
    \includegraphics[width=1\linewidth]{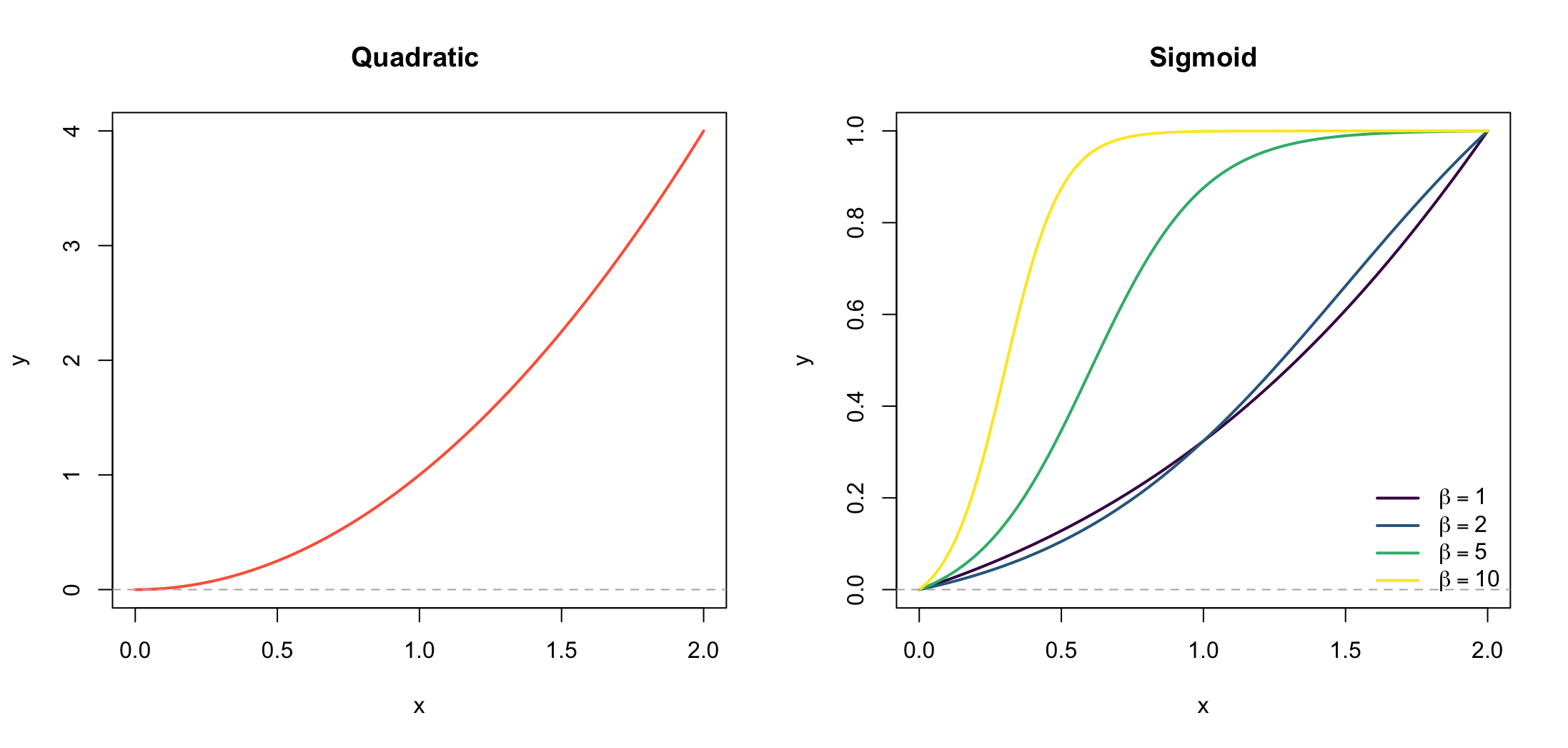}
    \caption{Decay curves for the quadratic (left)  and sigmoid (right) models.}
    \label{fig: curves}
\end{figure}
Figure~\ref{fig: curves} shows the quadratic curve and the sigmoid curves with different parameters. Except for the function applied to the distance, the rest of the sigmoid decay is similar to the quadratic decay, which can be written as
\begin{equation*} 
    \boldsymbol{\Xi} = \alpha_{1} \mathbf{J} - \alpha_{2} \mathbf{D}(\beta) + \alpha_{3} \mathbf{I},
\end{equation*}
where $\mathbf{D}(\beta)$ is the only different part. With the distance $d_{ij}$ measured on the coordinate system $\mathcal{C}$, the element of $\mathbf{D}(\beta)$ at the $i$th row and the $j$th column is $h(d_{ij})$. Hence, the element-wise version of the sigmoid decay is 
\begin{equation*}
    \xi_{ij} = \begin{cases}
  \alpha_1 - \alpha_2 h(d_{ij}) & \text{ if } i \ne j, \\
  \alpha_1 + \alpha_3 & \text{ if } i=j,
\end{cases}
\end{equation*}
where $\xi_{ij}$ is the element of $\boldsymbol{\Xi}$ at $(i,j)$. Three simulated $100 \times 100$ matrices are demonstrated in Figure~\ref{fig: simmaps} to show the effect on the spatial pattern of the QD and SD covariance structure. From the figure, we can see that the QD simulation image gradually decreases from the lower left to the upper right without any bulge. The SD-simulated image shows multiple obvious small bulges, and each pixel exhibits stronger local spatial positive correlation.

\subsection{Model Specification}
% Spatially Constrained Gaussian Mixture Mode
% The probability density function of a $G$-component spatially constrained Gaussian mixture model is defined as
% \begin{equation} \label{eq: scgmm}
%      f(\mathbf{x}\mid \boldsymbol{\vartheta}) = \sum_{g=1}^G \pi_g\phi(\mathbf{x}\mid\boldsymbol{\mu}_g,\boldsymbol{\Xi}_g).
% \end{equation}
% where 

The probability density of a $G$-component Gaussian mixture model with the sigmoid decay covariance structure (SpatGMM) is
\begin{equation} \label{eq: model}
     f(\mathbf{x}\mid \boldsymbol{\vartheta}) = \sum_{g=1}^G \pi_g\phi(\mathbf{x}\mid\boldsymbol{\mu}_g,\boldsymbol{\Xi}_g),
\end{equation}
where $\pi_g>0$, such that $\sum_{g=1}^G\pi_g = 1$, is the $g$th mixing proportion, $\boldsymbol{\mu}_g$ is the $g$th group mean vector, $\boldsymbol{\Xi}_g$ is the $g$th group covariance matrix, $\phi$ is the $p$-dimensional multivariate Gaussian density function, and $\boldsymbol{\vartheta} = \left\{\pi_1,\dots,\pi_G,\boldsymbol{\mu}_1,\dots,\boldsymbol{\mu}_G,\boldsymbol{\Xi}_1,\dots,\boldsymbol{\Xi}_G\right\}$ is the parameter space. With the coordinate system $\mathcal{C}$ and the sigmoid function $h(d)$, $\boldsymbol{\Xi}_g$ can be expressed as
\begin{equation*} 
    \boldsymbol{\Xi}_g = \alpha_{1g} \mathbf{J} - \alpha_{2g} \mathbf{D}(\beta_g) + \alpha_{3g} \mathbf{I},
\end{equation*}
where $\alpha_{1g}, \alpha_{2g}, \alpha_{3g} \in \mathbb{R}_+$, the $(i,j)$ element in $\mathbf{D}(\beta_g)$ is the value of $h(d_{ij})$ evaluated on distance $d_{ij}$, which is
\begin{equation}
    D_{ij} = h(d_{ij}) = a\left( \frac{1}{1+e^{-\beta_g d_{ij} + 3}} - s \right),
\end{equation}
where $a$ and $s$ have the same definition as above, and $\beta_g$ is the sigmoid parameter.

% Suppose a set of \textcolor{red}{i.i.d.} $p_1\times p_2\times \cdots \times p_m$ random tensor $\mathscr{X}_i$ , for $i=1,2,\dots, N$, along with a shared coordinate system $\mathcal{C}$, have the $p$-dimensional vectorizations $\mathbf{X}_i$ arising from a $G$-component spatially constrained Gaussian mixture model, where $p = \prod_{i=1}^m p_i$. Let $\mathbf{z}_i = (z_{i1},\dots,z_{iG})$ be the latent membership, where $z_{ig} = 1$ means $\mathbf{X}_i$ belong the $g$th component. The spatially constrained Gaussian mixture model assumes that
% \begin{equation*}
%     \mathbf{X}_i\mid z_{ig} = 1 \sim \mathcal{N}(\boldsymbol{\mu}_g,\boldsymbol{\Xi}_g),
% \end{equation*}
% where $\boldsymbol{\Xi}_g$ is the linear spatial covariance matrix. With the Euclidean distance $d_{ij}$ evaluated on the coordinate system $\mathcal{C}$ and the function $h(x)$, $\boldsymbol{\Xi}_g$ can be expressed as
% \begin{equation*} 
%     \boldsymbol{\Xi}_g = \alpha_{1g} \mathbf{J} - \alpha_{2g} \mathbf{D}(\beta_{1g},\beta_{2g},\dots,\beta_{qg}) + \alpha_{3g} \mathbf{I},
% \end{equation*}
% where $\alpha_{1g}, \alpha_{2g}, \alpha_{3g} \in \mathbb{R}_+$, each element in $\mathbf{D}(\beta_{1g},\beta_{2g},\dots,\beta_{qg})$ is the value of $h(x)$ evaluated on distance $d_{ij}$, which is
% \begin{equation}
%     D_{ij} = h(d_{ij}) = \sum_{j=1}^{q+r+1} \beta_{jg}I_j(d_{ij} \mid q, \mathbf{t}),
% \end{equation}
% where $\beta_{jg} \geq 0$, $\sum_{j=1}^M \beta_{jg} = 1$, $q$ is the degree of the I-splines, and $\mathbf{t}$ is a collection knots.

\subsection{Parameter Estimation}

For parameter estimation, due to the linear spatial covariance parameter $\boldsymbol{\alpha}_g$ and the sigmoid parameter $\beta_g$, the ordinary EM algorithm for the GMM is not sufficient. As mentioned in Section~\ref{sec: linear spatial covariance structure}, the GLS estimates are also MLEs, which allows us to embed the GLS estimator into the EM algorithm. Therefore, a variant of the EM algorithm is employed to perform parameter estimation for the proposed model. Suppose we have the observations $\mathbf{x}_1,\mathbf{x}_2,\dots,\mathbf{x}_N$, which share a common coordinate system $\mathcal{C}$. Considering the corresponding latent component membership $\mathbf{z}_1, \mathbf{z}_2, \dots, \mathbf{z}_N$, the complete-data log-likelihood function is
\begin{equation} \label{eq: complete log-likelihood}
    \ell_c(\boldsymbol{\vartheta}) = \sum_{i=1}^N \sum_{g=1}^G z_{ig}[\log \pi_g + \log \phi(\mathbf{x}_i \mid \boldsymbol{\mu}_g,\boldsymbol{\Xi}_g)].
\end{equation}
 
\paragraph{E-step:}
In the E-step, the conditional expectation of \eqref{eq: complete log-likelihood} is calculated and used for estimating the latent membership $\mathbf{z}$. Given the estimates of distribution parameters, the update equation of $\hat{\mathbf{z}}$ is 
\begin{equation*}
    \hat{z}_{ig} = \frac{\hat{\pi}_g \phi(\mathbf{x}_i \mid \hat{\boldsymbol{\mu}}_g,\hat{\boldsymbol{\Xi}}_g)}{\sum_{h=1}^G \hat{\pi}_h\phi(\mathbf{x}_i \mid \hat{\boldsymbol{\mu}}_h,\hat{\boldsymbol{\Xi}}_h)},
\end{equation*}
for $i=1,\dots,N$ and $g=1,\dots,G$. 

% the latent membership $\mathbf{z}$ is estimated with conditioning on the estimated parameters $\hat{\pi}_g$, $\hat{\boldsymbol{\mu}}_g$, and $\hat{\boldsymbol{\Xi}}_g$. Hence, the complete log-likelihood in the E-step is
% \begin{equation*}
%     \ell_c^{(1)}(\boldsymbol{\vartheta}) = C - \frac{1}{2} \sum_{i=1}^N \sum_{g=1}^G {z}_{ig} \operatorname{tr}\left\{(\mathbf{x}_i-\hat{\boldsymbol{\mu}}_g)(\mathbf{x}_i-\hat{\boldsymbol{\mu}}_g)^\prime
%     \hat{\boldsymbol{\Xi}}_g^{-1}\right\},
% \end{equation*}
% where $C$ is a constant with respect to $z_{ig}$. Therefore, the component membership $z_{ig}$ can be updated via 
% \begin{equation*}
%     \hat{z}_{ig} = \frac{\hat{\pi}_g \phi(\mathbf{x}_i \mid \hat{\boldsymbol{\mu}}_g,\hat{\boldsymbol{\Xi}}_g)}{\sum_{h=1}^G \hat{\pi}_h\phi(\mathbf{x}_i \mid \hat{\boldsymbol{\mu}}_h,\hat{\boldsymbol{\Xi}}_h)},
% \end{equation*}
% for $i=1,\dots,N$ and $g=1,\dots,G$. 

\paragraph{M-step 1:} In the first M-step, the mixing proportions $\boldsymbol{\pi}_g$, and the component means $\boldsymbol{\mu}_g$ are estimated first with conditioning on the estimated latent membership $\hat{z}_{ig}$. By substituting the estimated latent membership into the complete log-likelihood, the complete-data log-likelihood is 
\begin{equation*}
    \ell_c^{(2)}(\boldsymbol{\vartheta}) = C + \sum_{g=1}^G N_g \log \pi_g - \frac{1}{2} \sum_{g=1}^G \hat{z}_{ig} \operatorname{tr}\left\{(\mathbf{x}_i-\boldsymbol{\mu}_g)(\mathbf{x}_i-\boldsymbol{\mu}_g)^\prime
    \hat{\boldsymbol{\Xi}}_g^{-1}\right\},
\end{equation*}
where $N_g = \sum_{i=1}^N \hat{z}_{ig}$. By calculating the derivative of the log-likelihood function and setting it equal to zero, the update equations for $\hat{\pi}_g$ and $\hat{\boldsymbol{\mu}}_g$ separately are
\begin{equation*}
    \hat{\pi}_g = \frac{N_g}{N}, \quad  \quad\hat{\boldsymbol{\mu}}_g = \frac{1}{N_g} \sum_{i=1}^N \hat{z}_{ig}\mathbf{x}_i.
\end{equation*}

\paragraph{M-step 2:}
In the second M-step, the marginal log-likelihood with respect to $\boldsymbol{\Xi}_g$ is 
\begin{equation} \label{eq: single system loglikelihood3}
    \begin{aligned}
        \ell_c^{(3)}(\boldsymbol{\vartheta}) = C &- \frac{1}{2}\sum_{g=1}^G N_g\log|\boldsymbol{\Xi}_g| - \frac{1}{2} \sum_{g=1}^G N_g \operatorname{tr}\left\{\mathbf{S}_g
        {\boldsymbol{\Xi}}_g^{-1}\right\} \\
        = C &- \frac{1}{2}\sum_{g=1}^G N_g\log|\alpha_{1g}\mathbf{J} - \alpha_{2g}\mathbf{D}(\beta_g) + \alpha_{3g}\mathbf{I}| \\
        &  - \frac{1}{2} \sum_{g=1}^G N_g \operatorname{tr}\left\{\mathbf{S}_g
        (\alpha_{1g}\mathbf{J} - \alpha_{2g}\mathbf{D}(\beta_g) + \alpha_{3g}\mathbf{I})^{-1}\right\}
    \end{aligned}
\end{equation}
where $C$ is a constant in terms of $\boldsymbol{\Xi}_g$, and
\begin{equation*}
    \mathbf{S}_g = \frac{1}{N_g}\sum_{i=1}^N \hat{z}_{ig}(\mathbf{x}_i-\boldsymbol{\mu}_g)(\mathbf{x}_i-\boldsymbol{\mu}_g)^\prime.
\end{equation*}

Because the GLS estimators can provide us with MLE solutions, maximizing \eqref{eq: single system loglikelihood3} is equivalent to minimizing
\begin{equation}\label{eq: glse_g}
    g(\boldsymbol{\alpha}) = \frac{1}{2} (\mathbf{s}_g-\boldsymbol{\Delta}_g\boldsymbol{\alpha})^\prime(\mathbf{V}^*\otimes\mathbf{V}^*)(\mathbf{s}_g-\boldsymbol{\Delta}_g\boldsymbol{\alpha}),
\end{equation}
where $\mathbf{s}_g$ is the vectorization of $\mathbf{S}_g$, $\boldsymbol{\Delta}_g$ is the design matrix which columns are $\operatorname{vec}(\mathbf{J}), \operatorname{vec}(\mathbf{D}(\hat{\beta}_g)),\operatorname{vec}(\mathbf{I})$, and $\mathbf{V}^* = \left\{\hat{\boldsymbol{\Xi}}^*_g\right\}^{-1}$, where $\hat{\boldsymbol{\Xi}}^*_g$ is the estimated covariance matrix calculated in the previous EM iteration. By setting the first derivative of \eqref{eq: glse_g} equal to zero, we get the update equation
\begin{equation} \label{eq: alpha g}
    \hat{\boldsymbol{\alpha}}_g = \left\{\boldsymbol{\Delta}_g^\prime(\mathbf{V}^*\otimes \mathbf{V}^*)\boldsymbol{\Delta}_g\right\}^{-1}\boldsymbol{\Delta}_g^\prime\operatorname{vec}(\mathbf{V}^*\mathbf{S}_g\mathbf{V}^*).
\end{equation}

\paragraph{M-step 3:} In the last M-step, with $\hat{\boldsymbol{\alpha}}_g$, to estimate the sigmoid parameter $\beta_g$, we minimize 
\begin{equation}\label{eq: obj for beta}
    F_g = \log |\boldsymbol{\Xi}_g| - \log |\mathbf{S}_g| + \operatorname{tr}\left\{\mathbf{S}_g
    {\boldsymbol{\Xi}}_g^{-1}\right\} -p,
\end{equation}
where $\boldsymbol{\Xi}_g = \hat{\alpha}_{1g}\mathbf{J} - \hat{\alpha}_{2g}\mathbf{D}(\beta_g) + \hat{\alpha}_{3g}\mathbf{I}$, with respect to $\beta_g$. Minimizing \eqref{eq: obj for beta} is equivalent to maximizing the likelihood function, so it aligns with the EM algorithm framework. Because of the complexity of the sigmoid function \eqref{eq: sigmoid function}, \eqref{eq: obj for beta} is difficult to optimize. Hence, we use \pkg{optimize} in \proglang{R} to compute the solution.

\section{Numerical Experiments and Analysis} \label{Apps}

Three simulation studies are presented to evaluate the performance of the proposed model. Furthermore, the methodology is applied to Raman spectroscopy (RS) data to demonstrate its capacity in practical scenarios. For all simulation studies and the real data application in this section, the coordinate system is defined by the subscripts of the spatial system. But it is also possible to specify a particular coordinate system. In addition, as discussed in Section~\ref{sec: cs}, we normalize the Euclidean distance evaluated on the given coordinate system to be in the range $[0,2]$. In the following simulation studies and real data analysis, the normalization is always performed. The classification accuracy is evaluated using the adjusted Rand index \citep[ARI;][]{hubert1985comparing}, which is an improvement over the Rand index \citep[RI;][]{rand1971objective}. 

\subsection{Simulation Studies}
Three Monte Carlo simulation studies are conducted to evaluate different aspects of the proposed mixture models. The first simulation design is conducted to demonstrate the effectiveness of spatial parameters $\boldsymbol{\alpha}_g$ and $\beta_g$. In different groups, only the spatial parameters being tested will vary, while all other parameters will remain the same. The second numerical experiment aims to demonstrate the capacity of the Bayesian information criterion \citep[BIC;][]{schwarz78} to detect the correct specification of the number of components, $G$. In the third simulation design, MCLUST \citep{scrucca2016mclust}, PGMM, and MMN are compared with the proposed approach to demonstrate its superior performance on spatial data.

\subsubsection{Simulation Design I - Parameter Recovery} \label{sec: sim1}
\begin{figure}[ht]
    \centering
    \includegraphics[width=1\linewidth]{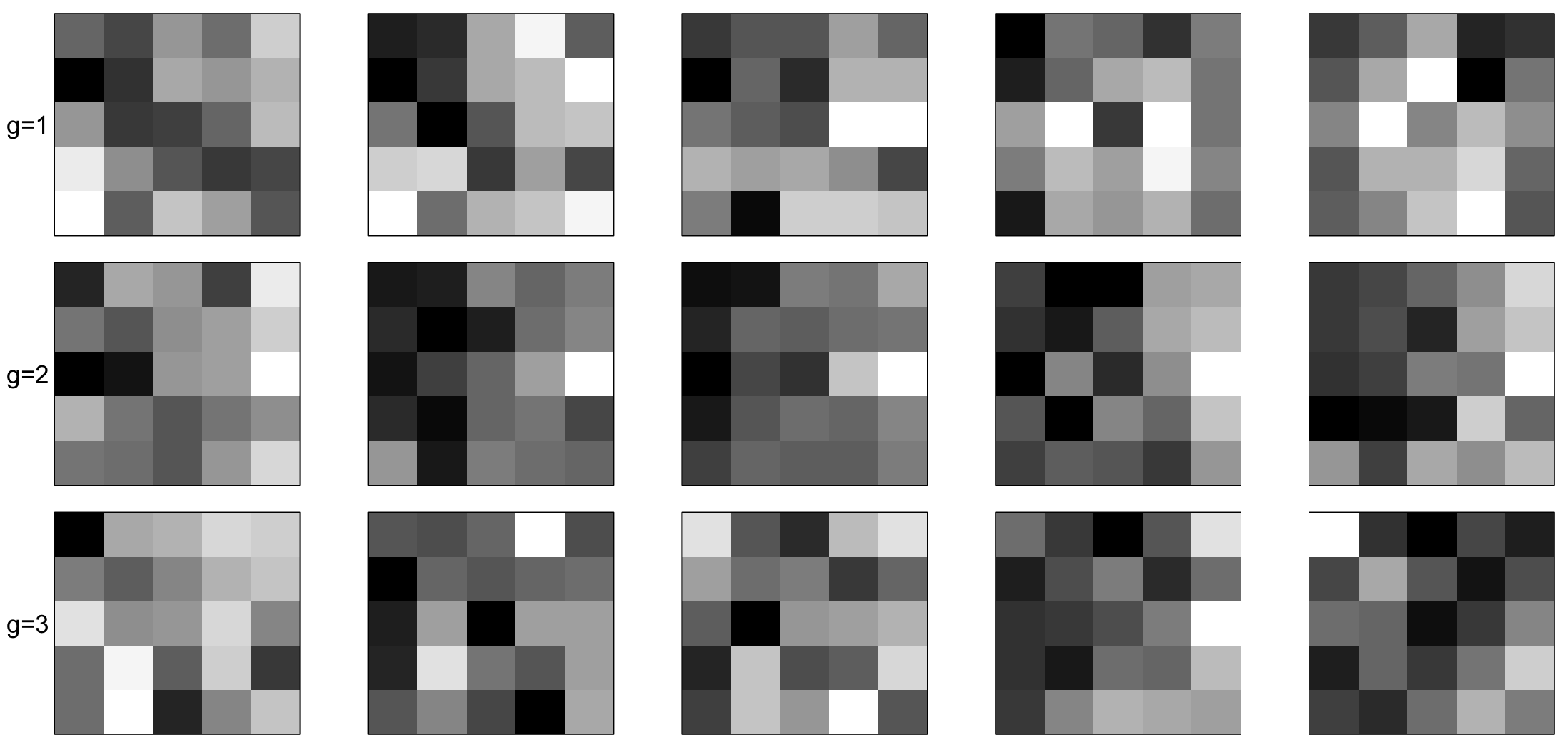}
    \caption{Visualization of the first observed tensor from each of the three groups in simulation design I.}
    \label{fig: viz of gen tensors}
\end{figure}

In the first setting, the number of groups is $G=3$, with the mixing proportions $(\pi_1,\pi_2,\pi_3) = (0.2,0.3,0.5)$. Each generated observation is a tensor of order three, with dimensions $5 \times 5 \times 5$ and all the group means are tensors of $0$s. The setting of spatial parameters $\boldsymbol{\alpha}_g$ and $\beta_g$ is shown in Table~\ref{tab: setting1 cov param}. Aside from $\boldsymbol{\alpha}_g$ and $\beta_g$, all the means in each group are shared to be the same. $50$ datasets of size $N=1000$ are generated from the proposed model. Figure~\ref{fig: viz of gen tensors} shows the visualization of the generated tensors from the three groups, which are difficult to classify with only the naked eye.

\begin{table}[ht]
\centering
\caption{True spatial parameters associated with each group in simulation design I.}
% temporarily reduce column separation
{\setlength{\tabcolsep}{6pt}% default is 6pt
 \renewcommand{\arraystretch}{1}% optional: tighten row spacing
 \begin{tabular}{c|ccc|c}
  \hline
  Group 
    & $\alpha_{1g}$ & $\alpha_{2g}$ & $\alpha_{3g}$ 
    & $\beta_{g}$  \\
  \hline
  1 & 4 & 3 & 2 & 4 \\
  2 & 2 & 1 & 1 & 4 \\
  3 & 4 & 3 & 2 & 10 \\
  \hline
 \end{tabular}
}
\label{tab: setting1 cov param}
\end{table}

The result of parameter recovery is illustrated in Table~\ref{tab: estimated alpha}, which is accurate with acceptable standard errors. With respect to the clustering performance, the average of ARI for each repeat is $0.98$.

\begin{table}[ht]
\centering
\caption{Estimated means and standard errors (Se) of $\hat{\boldsymbol{\alpha}}_g$ and $\hat{\beta}_g$ for simulation design I.}
\setlength{\tabcolsep}{6pt}
\begin{tabular}{
  c
  S[table-format=1.2] S[table-format=1.3]
  S[table-format=1.2] S[table-format=1.3]
  S[table-format=1.2] S[table-format=1.3]
  S[table-format=2.2] S[table-format=1.3]
}
\toprule
Group
  & \multicolumn{2}{c}{\(\hat{\alpha}_{1g}\)}
  & \multicolumn{2}{c}{\(\hat{\alpha}_{2g}\)}
  & \multicolumn{2}{c}{\(\hat{\alpha}_{3g}\)}
  & \multicolumn{2}{c}{\(\hat{\beta}_g\)} \\
\cmidrule(lr){2-3}\cmidrule(lr){4-5}\cmidrule(lr){6-7}\cmidrule(lr){8-9}
  & {Mean} & {Se}
  & {Mean} & {Se}
  & {Mean} & {Se}
  & {Mean} & {Se} \\
\midrule
1 & 3.92  & 0.133 & 2.94  & 0.130 & 1.97  & 0.051 &  4.00 & 1.848 \\
2 & 2.02  & 0.113 & 1.03  & 0.120 & 1.01  & 0.047 &  4.00 & 0.908 \\
3 & 3.99  & 0.111 & 2.99  & 0.061 & 2.00  & 0.034 & 10.01 & 0.108 \\
\bottomrule
\end{tabular}
\label{tab: estimated alpha}
\end{table}

\subsubsection{Simulation Design II - Model Selection} \label{sec: sim2}
In this study, to detect the capacity of BIC to select the correct number of components, we set $4$ groups with the dimensionality $5 \times 5 \times 5$, still with the mixing proportion $(\pi_1,\pi_2,\pi_3) = (0.2,0.3,0.5)$. For the location tensors, the four groups have different settings, which are the tensors of $0$, $5$, $10$, and $-5$. The setting of the covariance parameter is shown in Table~\ref{tab: setting2 cov param}.
\begin{table}[ht]
\centering
\caption{True spatial parameters associated with each group in simulation design II.}
% temporarily reduce column separation
{\setlength{\tabcolsep}{6pt}% default is 6pt
 \renewcommand{\arraystretch}{1}% optional: tighten row spacing
 \begin{tabular}{c|ccc|c}
  \hline
  Group 
    & $\alpha_{1g}$ & $\alpha_{2g}$ & $\alpha_{3g}$ 
    & $\beta_{g}$  \\
  \hline
  1 & 4 & 3 & 2 & 4 \\
  2 & 2 & 1 & 1 & 4 \\
  3 & 4 & 3 & 2 & 10 \\
  4 & 1 & 1 & 1 & 10 \\
  \hline
 \end{tabular}
}
\label{tab: setting2 cov param}
\end{table}
Given the parameter setting, we generate $50$ datasets from our proposed model, each of size $1000$. Table~\ref{tab: frequency_table} shows the best model setting with the highest BIC in all $50$ repeats. 

\begin{table}[!ht]
\centering
\caption{Model selection performance via the BIC for SpatGMM for simulation design II. True number of components is 4.}
\label{tab: frequency_table}
\begin{tabular}{c|ccccc}
\toprule
  & $G=2$ & $G=3$ & $G=4$ & $G=5$ & $G=6$\\
\midrule
       Frequency & 0 & 0 & 38 & 12 & 0\\
\bottomrule
\end{tabular}
\end{table}

\subsubsection{Simulation Design III} \label{sec: sim3}
In the third simulation, we compare the proposed model to MCLUST, PGMM, and MMN. $30$ datasets are generated, which follow the probability density \eqref{eq: model} with $G=3$. In each dataset, the shape of the observations is $10 \times 10$, but the setting of the distribution parameter is the same as in Section~\ref{sec: sim1}. For MCLUST, according to the BIC score, the `EEE' model is selected, but the average ARI of $30$ repeats is $0$, which indicates that MCLUST barely works for the simulated datasets. In terms of the covariance structure of PGMM, since our spatial covariance structure is isotropic, it is reasonable to choose the `CUU' model, which is also what the BIC suggests. Furthermore, the range of the potential number of factors is $\{9, 10, 11, 12, 13 \}$, and we use the BIC to select the best model. 
\begin{table}[ht]
\centering
\caption{Clustering performance of four competing models for simulation design III. }
\label{tab: compare1}
\begin{tabular}{ccccc}
\toprule
  & SpatGMM & MCLUST & PGMM & MMN\\
\midrule
 Average ARI & 0.98 & 0.00 & 0.47 & 0.46\\
 Standard deviation of ARI & 0.0066 & 0.0014 & 0.0316 & 0.0412 \\
\bottomrule
\end{tabular}
\end{table}

\begin{table}[ht]
\centering
\caption{BIC of four competing models for simulation design III. }
\label{tab: compare2}
\begin{tabular}{r|rrrr}
\toprule
  & SpatGMM & MCLUST & PGMM & MMN\\
\midrule
 Average BIC & -181,898 & -405,012 & -381,437 & -382,563\\
 Standard deviation of BIC & 1,541 & 1,299 & 1,572 & 1,546 \\
\bottomrule
\end{tabular}
\end{table}
Based on Tables~\ref {tab: compare1} and~\ref {tab: compare2}, we see that for data with spatial covariance, the proposed model is the most appropriate choice.

\subsection{Real Application}

\begin{figure}
    \centering
    \includegraphics[width=1\linewidth]{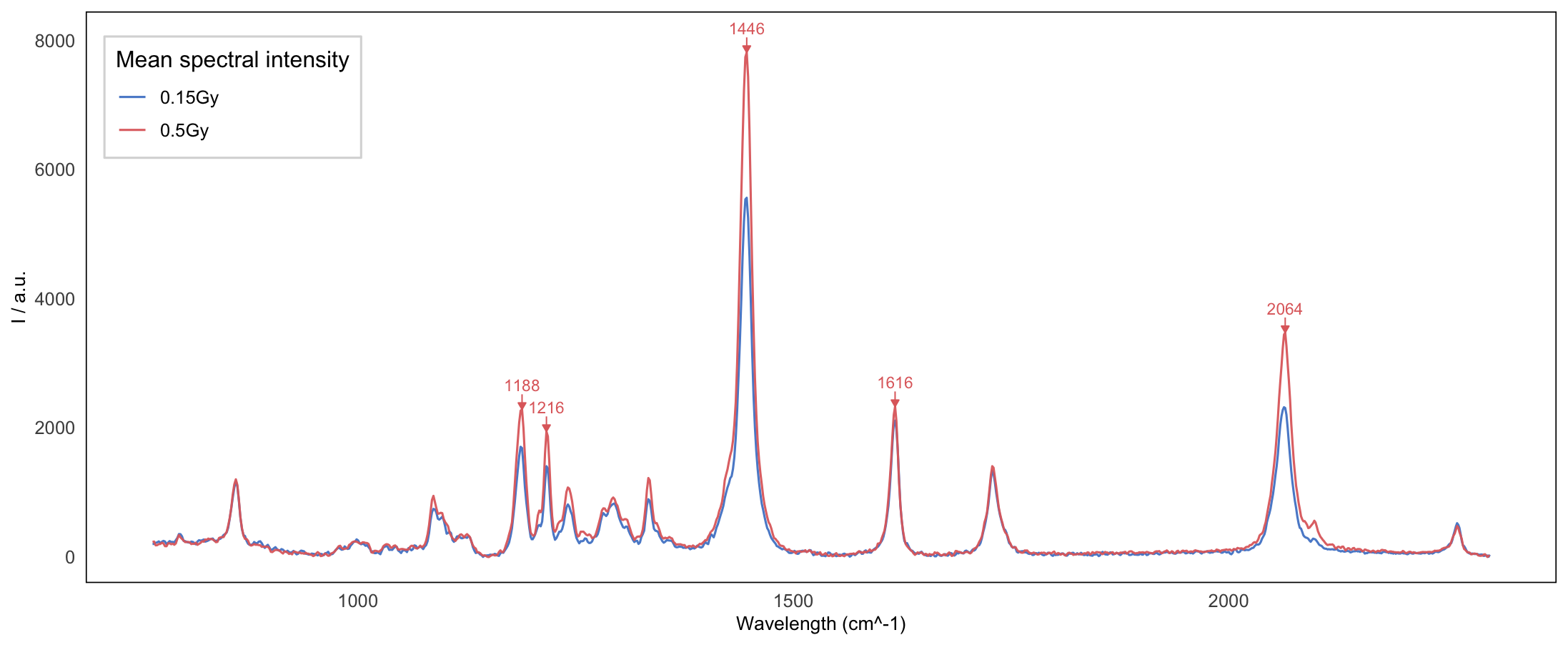}
    \caption{Raman spectra of dosimetric film at two different dosage levels.}
    \label{fig: raman spec}
\end{figure}

Following simulated experiments, the performance of the spatially constrained Gaussian mixture model(SpatGMM) is evaluated using Raman spectroscopy data. Raman spectroscopy (RS) measures relative molecular abundances and dynamics. Using a microscope platform, RS can provide spatial resolution on the order of micrometers \citep{mulvaney2000raman,das2011raman,orlando2021comprehensive}. The Raman spectra analyzed in this study were obtained from measurements on a radiochromic dosimeter called EBT-3 film \citep{mcnairn2025exploring}. Upon exposure to ionizing radiation, the active layer between two matte polyester layers undergoes polymerization, resulting in a stable color change proportional to the absorbed dose \citep{borca2013dosimetric}. RS provides a non-invasive, high-resolution approach for detecting polymerization in EBT-3 films by monitoring peaks associated with dose-dependent changes, as shown in Figure~\ref{fig: raman spec}.

\begin{figure}
    \centering
    \includegraphics[width=1\linewidth]{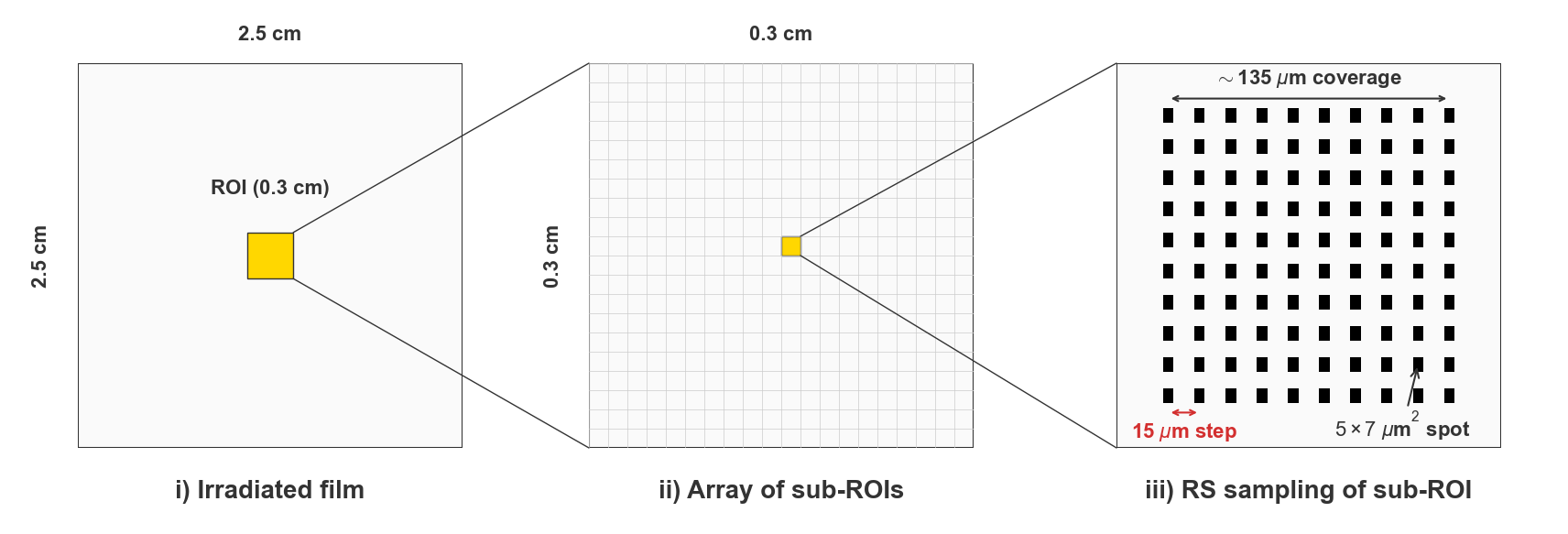}
    \caption{Schematic representation of the RS sampling protocol. (i) Macroscopic view of the irradiated EBT-3 film ($2.5 \times 2.5$ cm). (ii) The ROI comprises an array of 400 separate sub-ROIs (grids) arranged in a $20 \times 20$ pattern. (iii) Microscopic view of a sub-ROI, illustrating the $10 \times 10$ point-scan geometry.}
    \label{fig: Figure_1}
\end{figure}

\begin{figure}
    \centering
    \includegraphics[width=0.75\linewidth]{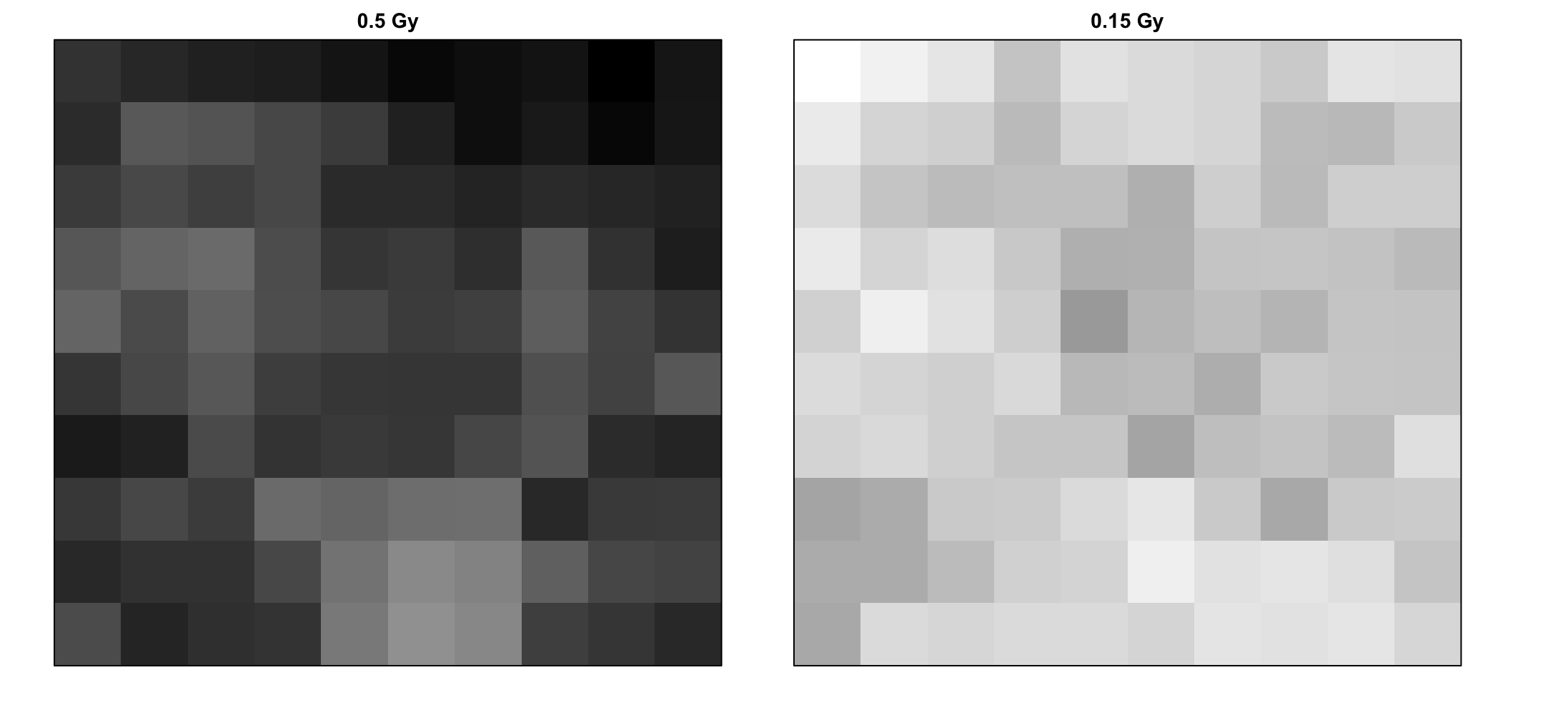}
    \caption{Visualization of a Raman peak at $2064$ $\text{cm}^{-1}$ measured on a $10 \times 10$ grid across two dosimetric films exposed to different radiation levels (0.5 Gy left, 0.15 Gy right).}
    \label{fig: examples of tensors two dose levels}
\end{figure}

Data were collected from a 0.3 cm square region of interest (ROI) containing an array of 400 distinct sub-ROIs, as illustrated in Figure~\ref{fig: Figure_1}. In our analysis, each sub-ROI is treated as a matrix-variate observation. Each observation consists of a $10 \times 10$ point-scan grid acquired with a $15 \mu$m step size and a $5 \times 7 \mu$m$^2$ spot size, providing a spatial coverage of approximately $135 \times 135 \mu$m$^2$.
The dataset comprises 800 such observations (totaling 80,000 spectra) collected from two films exposed to radiation doses of $0.15$ Gy and $0.50$ Gy, respectively. The spectra were recorded over a wavenumber range of $764$ to $2300$ $\text{cm}^{-1}$. For the spatial analysis, we selected the dose-dependent peak at $2064$ $\text{cm}^{-1}$, as highlighted in Figure~\ref{fig: raman spec}. Representative tensor observations for both dose levels are visualized in Figure~\ref{fig: examples of tensors two dose levels}.

\begin{figure}[ht]
    \centering
    \includegraphics[width=0.5\linewidth]{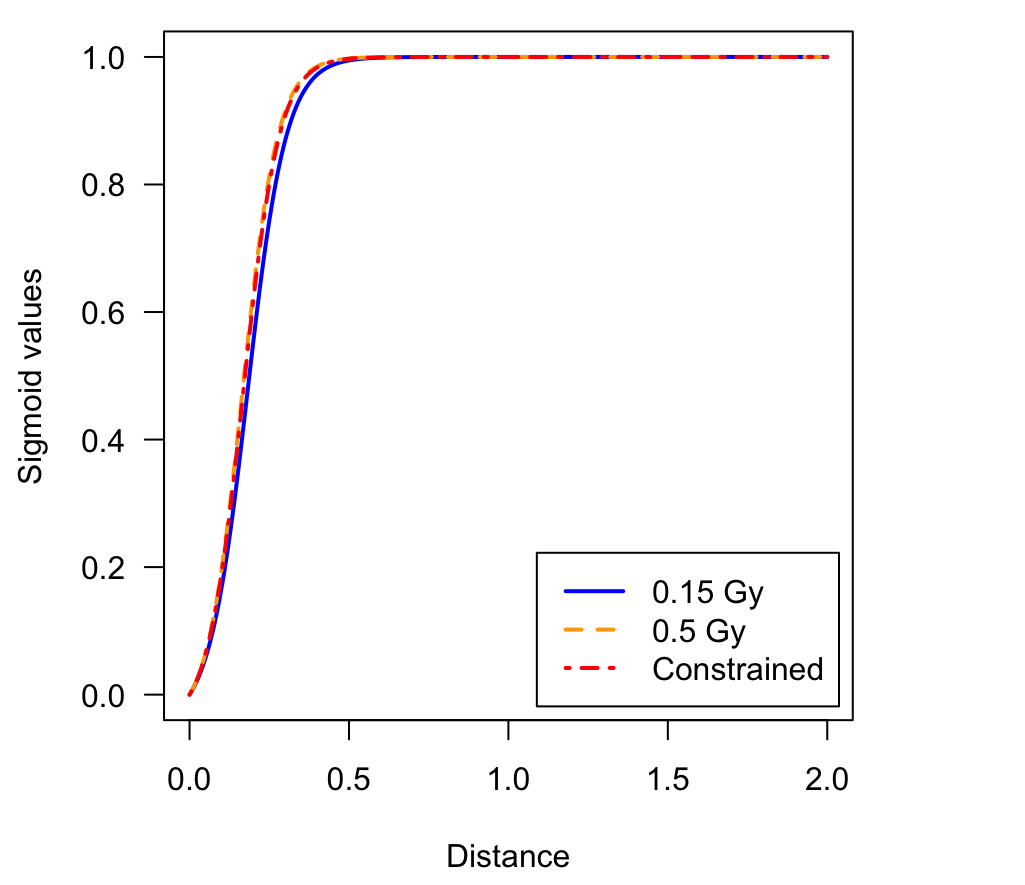}
    \caption{Estimated sigmoid functions in application to dosimetric films data.}
    \label{fig: raman sigmoid}
\end{figure}

% \begin{figure}[ht]
%     \centering
%     \includegraphics[width=0.75\linewidth]{images/Raman observation.png}
%     \caption{Raman maps of two dose levels at $2064$ $\text{cm}^{-1}$}
%     \label{fig: raman films of two dose levels}
% \end{figure}
% Prior to applying the spatially constrained Gaussian mixture model, the data are normalized.
The estimated spatial parameters for the two groups corresponding to the two absorbed dose magnitudes are illustrated in Table~\ref{tab: real data cov param}. Figure~\ref{fig: raman sigmoid} displays the sigmoid function with these estimated parameters.
In Figure~\ref{fig: examples of tensors two dose levels}, we can easily see that the two groups are very different from each other in magnitude. Therefore, clustering the spectra on the basis of their dose level is a straightforward problem that can be effectively solved by most clustering techniques. Hence, we centralize the data within each dosage level. The proposed model can still accurately capture the covariance and group structure of the centered data, but MCLUST and MMN no longer work. Regarding PGMM, we set the number of factors to be in the range of 4 to 16 and use the BIC to select the best model. According to the BIC, the `CUC' model with $16$ factors is selected. Although the PGMM model achieves good clustering accuracy, as shown in Table~\ref{tab: compare3}, it has a lower BIC score. 
\begin{table}[ht]
\centering
\caption{Comparison of performance between PGMM and SpatGMM on group-centered dosimetric film data.}
\label{tab: compare3}
\begin{tabular}{r|rr}
\toprule
  & ARI & BIC\\
\midrule
 SpatGMM & 0.926 & -918,876  \\
 PGMM & 0.883 & -1,081,612 \\
\bottomrule
\end{tabular}
\end{table}

In addition to the clustering results, we also focus on the spatial pattern of the Raman maps. According to the estimated spatial parameters, there is a temptation to think of these parameters as `similar'. To assess whether spatial patterns differ significantly, a constrained model is implemented in which all spatial covariance parameters are shared between groups, which is $\boldsymbol{\alpha}_1 = \boldsymbol{\alpha}_2$ and $\beta_1 = \beta_2$. The estimated spatial parameters of the constrained model are also listed in Table~\ref{tab: real data cov param} for comparison. Table~\ref{tab: real data cov param} also presents the BIC scores for both the unconstrained and constrained models. 
\begin{table}[ht]
\centering
\caption{Estimated spatial parameters for SpatGMM on dosimetric film data with individual parameters per group (first two rows) and parameters constrained across groups (third row).}
% temporarily reduce column separation
{\setlength{\tabcolsep}{6pt}% default is 6pt
 \renewcommand{\arraystretch}{1}% optional: tighten row spacing
 \begin{tabular}{c|rrr|r|r}
  \hline
  Group 
    & $\hat{\alpha}_{1g}$ & $\hat{\alpha}_{2g}$ & $\hat{\alpha}_{3g}$ 
    & $\hat{\beta}_{g}$  & BIC \\
  \hline
  0.5 Gy & 81,364 & 67,207 & 20,563 & 18.1 & \multirow{2}{5em}{\makebox[5em][r]{-919,001}}\\
  0.15 Gy & 33,484 & 19,415 & 8,958 & 16.5 & \\
  \hline
  Constrained & 57,584 & 43,448 & 14,623 & 17.8 & -930,558 \\
  \hline
 \end{tabular}
}
\label{tab: real data cov param}
\end{table}
As the BIC value for the unconstrained model is higher than that for the constrained model in Table~\ref{tab: real data cov param}, statistically, we can argue that the spatial polymerization patterns of EBT-3 films at the micron-scales level differ between these two radiation dosages.

\section{Summary} \label{sec:summary}

A finite Gaussian mixture model with spatial constraints is proposed for clustering tensor variate data that exhibits intrinsic spatial correlation. Given a coordinate system, the covariance matrices can be modeled as sigmoid function values of spatial proximity, which is the sigmoid decay. The proximity within the spatial system is measured by the Euclidean distance between the coordinates, which is then normalized to maintain consistency. The structure can first help us infer the sigmoid-like spatial correlation between elements in the spatial systems of different groups. Furthermore, it offers the finite mixture model substantial parsimony. No matter how large the dimensionality is, this structure can model the covariance matrix with only a constant number of free parameters, which is $5$. In addition, along with the GLS estimator, a variant of the EM algorithm is illustrated for parameter estimation.  

Although the sigmoid decay structure offers some benefits, there are still aspects of it that require improvement. First, even if the sigmoid decay provides more flexibility than the quadratic decay, it still can be restrictive in some cases, because the sigmoid assumption may also not hold water. It is general to assume that the spatial correlation decreases as the distance increases, but the relationship may not be perfectly sigmoid-like. Therefore, a more flexible framework is required. Second, under the current structure, we also assume the variance of each element in the spatial systems is the same. However, this is not the case in many instances, such as with pixel values in a collection of pictures. Third, the parameter estimation procedure for the sigmoid parameter is not stable. During the optimization procedure for estimating the sigmoid parameter, the corresponding covariance matrix may be singular, which will affect the calculation of the objective function. Moreover, the calculation of the inversion of the covariance matrix remains a problem, despite its well-structured nature. However, suppose the subscripts of the elements are used as the coordinate system. In that case, the covariance matrix is also a Toeplitz block Toeplitz matrix, which has a special trick for inversion calculations \citep{wax2003efficient}.

\bibliographystyle{apalike}
\bibliography{EntireLibrary}

\end{document}